\documentclass[aps,prb,reprint,twocolumn,tightenlines,showpacs,superscriptaddress,floatfix]{revtex4-2}
\usepackage{graphicx}% Include figure files
\usepackage{dcolumn}% Align table columns on decimal point
\usepackage{bm, amsmath,amssymb}% math
\usepackage{float}
\usepackage{url}
\usepackage{layout}
\usepackage{booktabs}
\usepackage{color}
\usepackage[hyperindex,breaklinks]{hyperref}

\begin{document}
\title{Universality classes of the Anderson transition in three-dimensional symmetry classes AIII, BDI, C, D and CI}

\author{Tong Wang}
%\thanks{}
\affiliation{International Center for Quantum Materials, School of Physics, Peking University, Beijing 100871, China}
\affiliation{Collaborative Innovation Center of Quantum Matter, Beijing 100871, China}

\author{Tomi Ohtsuki}
\affiliation{Physics Division, Sophia University, Chiyoda-ku, Tokyo 102-8554, Japan}

\author{Ryuichi Shindou}
\email{rshindou@pku.edu.cn}
\affiliation{International Center for Quantum Materials, School of Physics, Peking University, Beijing 100871, China}
\affiliation{Collaborative Innovation Center of Quantum Matter, Beijing 100871, China}

\date{\today}

\begin{abstract}
	We clarify universal critical properties of delocalization-localization transitions in three-dimensional (3D) unitary and orthogonal classes with particle-hole and/or chiral symmetries (classes AIII, BDI, D, C and CI). We first introduce tight-binding models on cubic lattice that belong to these five nonstandard symmetry classes respectively. Unlike the Bogoliubov-de Gennes  Hamiltonian for superconductors, all the five models have finite areas of Fermi surfaces in the momentum space in the clean limit. Thereby, the scaling theory of the Anderson transition guarantees the presence of the delocalization-localization transitions at finite disorder strength in these models. Based on this expectation, we carry out extensive transfer matrix calculations of the Lyapunov exponents for zero-energy eigenstates of the disordered tight-binding models with quasi-one-dimensional geometry. Near the Anderson transition point, the correlation length diverges with a universal critical exponent $\nu$. Using finite-size scaling analysis of the localization length, we determine the critical exponent of the transitions and scaling function of the (normalized) localization length in the three non-standard unitary symmetry classes: $\nu_{\text{AIII}}=1.06\pm0.02$,  $\nu_{\text{D}}=0.87\pm0.03$, and $\nu_{\text{C}}=0.996\pm0.012$. Our result of the class C is consistent with a previous study of classical network model [M. Ortuño et al, Phys. Rev. Lett. \textbf{102}, 070603 (2009)]. The critical exponents of the two non-standard orthogonal classes are estimated as $\nu_{\text{CI}}=1.17 \pm 0.02$ and $\nu_{\text{BDI}}=1.12 \pm 0.06$.
	Our result of the class CI is consistent with another previous study of a lattice model, while the exponent of the class BDI is at variance with the previous evaluation of nodal Dirac ring model [X. L. Luo \text{et al}, Phys. Rev. B. \textbf{101}, 020202(R) (2020)].
\end{abstract}

\maketitle

\section{Introduction}
The quantum interference of two counter-propagating waves form a standing wave that does not move 
in space, suppressing particle diffusions completely in random media. 
After Anderson's seminal proposal of the localization of electron wavefunctions in disordered solids~\cite{Anderson1958,Evers08,Anderson50years}, theoretical understandings of the localization phenomena have 
been elaborated by scaling theories~\cite{Wegner1976,Abrahams1979}, field theories~\cite{Efetov80,Hikami81,hikami81a,Gade91,Gade93},
numerical simulations~\cite{MacKinnon81,MacKinnon83,Pichard81,SlevinTomi1999}, and 
symmetry classifications of random matrices~\cite{Wigner51,Dyson62,Dyson62TFW,AZ1997PRB}. It is widely acknowledged that 
delocalization-localization (Anderson) transition occurs in a variety of 
physical systems, including many-body electronic 
systems \cite{MBLfirst,MBL_IsingChain,MBLreview}, Bose–Einstein 
condensates \cite{Billy08,ALinBEC1,ALinBEC2}, and classical optical~\cite{John87,Wiersma97,ALofLight,TransverseALofLight,Skipetrov15,Skipetrov18} and acoustic 
systems~\cite{Kirkpatrick85,WEAVER90,Hu08}. The Anderson transition is a continuous quantum phase transition. Like other second-order phase transitions in statistical physics, the quantum 
phase transitions are categorized by universality classes. Each universality class is 
characterized by universal critical exponent and scaling functions, the information of which is 
encoded into scaling properties of an effective field theory around its fixed points. 
It is widely believed that the universality class 
of the Anderson transition is determined only by symmetries of disordered Hamiltonian (random 
matrix) and spatial dimension of the system~\cite{Wegner1976,Abrahams1979}. 

\begin{table}[b]
\begin{tabular}{|c|c|c|c|c|c|c|}
\hline
 & class & TRS& PHS&CS&$\nu$&Ref.\\ \hline\hline
Unitary&A &No&No&No&$1.443\pm 0.003$ &\cite{Slevin16}\\ \hline
Orthogonal&AI&$1$&No&No& $1.572\pm 0.003$&\cite{Slevin18}\\ \hline
Symplectic&AII&$-1$&No&No &$1.37\pm 0.01$&\cite{Asada05}\\ \hline\hline
Chiral Unitary&AIII&No&No&Yes& $1.06 \pm 0.02$ & *\\ \hline
Chiral Orthogonal&BDI&$1$&$1$&Yes &$1.12\pm 0.06$ &*\\ \hline
Chiral Symplectic&CII&$-1$&$-1$&Yes &--&--\\ \hline\hline
   & D &No&1&No & $0.87\pm 0.03$ & *\\ \cline{2-7}
 BdG  & C &No&$-1$&No & $0.996\pm 0.012$& *\\ \cline{2-7}
   & DIII&$-1$&$1$&Yes& $1.1\pm 0.05$&\cite{roy2017}\\ \cline{2-7}
   & CI &1&$-1$&Yes &$1.17\pm 0.02$ & *\\ \hline\hline
\end{tabular}
\caption{Classification according to time reversal symmetry (TRS),  
particle-hole symmetry (PHS), chiral symmetry (CS), and their universal critical 
exponents $\nu$ in 3D Anderson transition.
TRS is defined as $\mathbb{T}\mathbb{H}^T \mathbb{T}^{-1} = \mathbb{H}$
and PHS as $ \mathbb{C}\mathbb{H}^T \mathbb{C}^{-1} = -\mathbb{H}$.
For TRS (PHS) 1 mean $\mathbb{T}^T=\mathbb{T}$ ($\mathbb{C}^T=\mathbb{C})$ whereas 
$-1$ means $\mathbb{T}^T=-\mathbb{T}$ $(\mathbb{C}^T=-\mathbb{C}) $.
 * indicates the values estimated in this paper,
whereas -- indicates yet to be determined.}
\label{tab:classification}
\end{table}

The classification of Hermitian random matrices leads to ten symmetry classes known as 
the Altland-Zirnbauer (AZ) classes.  They consist of  
standard Wigner-Dyson symmetry classes (A, AI, AII),
three chiral symmetry classes (AIII, BDI, CII)~\cite{Gade91,Gade93},
and four Bogoliubov-de Gennes (BdG) symmetry classes  (D, C, DIII, CI)~\cite{AZ1997PRB,AZ2005}. 
The random matrices and disordered Hamiltonian 
in the non-standard symmetry classes not only describe 
universal aspects of the localization of Bogoliubov quasiparticle 
wavefunctions in disordered superconductors~\cite{AZ1997PRB, AZ2005} 
but also they are closely related to the localization of classical 
waves in random dissipative media~\cite{kawabata21}, 
non-Hermitian disordered systems, which are attracting a lot of 
research interests during the last couple 
of years~\cite{xu16,tzortzakakis20,wang20,huang20a,huang20b,luo21a,luo21b,cao99}. 
As some will be demonstrated for the first time in this paper, 
three-dimensional (3D) models in each symmetry class show the  
delocalization-localization transition and its quantum criticality is 
characterized by its own universal scaling functions and 
the critical exponent. So far, the critical exponent of 
the 3D Anderson transition in seven out of the 
ten AZ classes have been studied numerically. 
They are the three Wigner-Dyson classes A, AI 
and AII \cite{Slevin16,Slevin18,Asada05}; the chiral orthogonal 
class BDI \cite{Luo2020PRB}; and three BdG classes 
C \cite{Chalker3dclassC}, CI \cite{Luo2020PRB}, and DIII \cite{roy2017}, 
while the other three (AIII, CII, D) remain to be clarified.
Numerical values of the critical exponent of these classes, including 
those evaluated in this paper, are listed in TABLE \ref{tab:classification}.

In this article, we clarify the universal critical exponents and scaling functions 
associated with the localization length in the 3D Anderson transition of the 
symmetry classes AIII, BDI, C, D and CI, using transfer matrix method. 
The chiral class AIII has topological phases classified by integer winding number 
in $d=1,3$ dimension(s). The study of the class AIII was rather established both 
numerically and analytically in 1D~\cite{HughesAIII2014,ProdanAIII2014,AIII_NLSM2017,BjornAIII2020,Hughes2dAIII2020}, 
but most of them focused on topological quantum critical point and few were 
directed to localization phenomena in 3D class AIII systems.
Class C and D in 3D are also less studied than in 2D, where there 
exists 2D spin and thermal quantum Hall insulator phases in these 
two symmetry classes, respectively. 
Low-energy Bogoliubov excitations in spin-triplet superconductors with 
broken time-reversal symmetry Sr$_2$RuO$_4$~\cite{Sr2RuO4_1998,Sr2RuO4_2006} 
and UPt$_3$~\cite{UPt3_2014} potentially belong to these 
two non-standard symmetry classes in 3D. 

In this paper, we propose two-band tight-binding models on the cubic lattice that 
belong to symmetry classes AIII, BDI C, D and CI respectively, which enable precise simulation studies of the 3D Anderson transition. 
Critical exponents of the Anderson transition in the 3D 
classes AIII and D are estimated for the first time, whose values are 
distinct from those of the other known symmetry classes.
Our estimates of exponents for the classes C and CI agree with previous
works of a classical network model 
\cite{Chalker3dclassC} and a lattice model \cite{Luo2020PRB}. 

\section{Unitary models with particle-hole or chiral symmetries}
\begin{figure*}[htb]
	\includegraphics[width=6.5in]{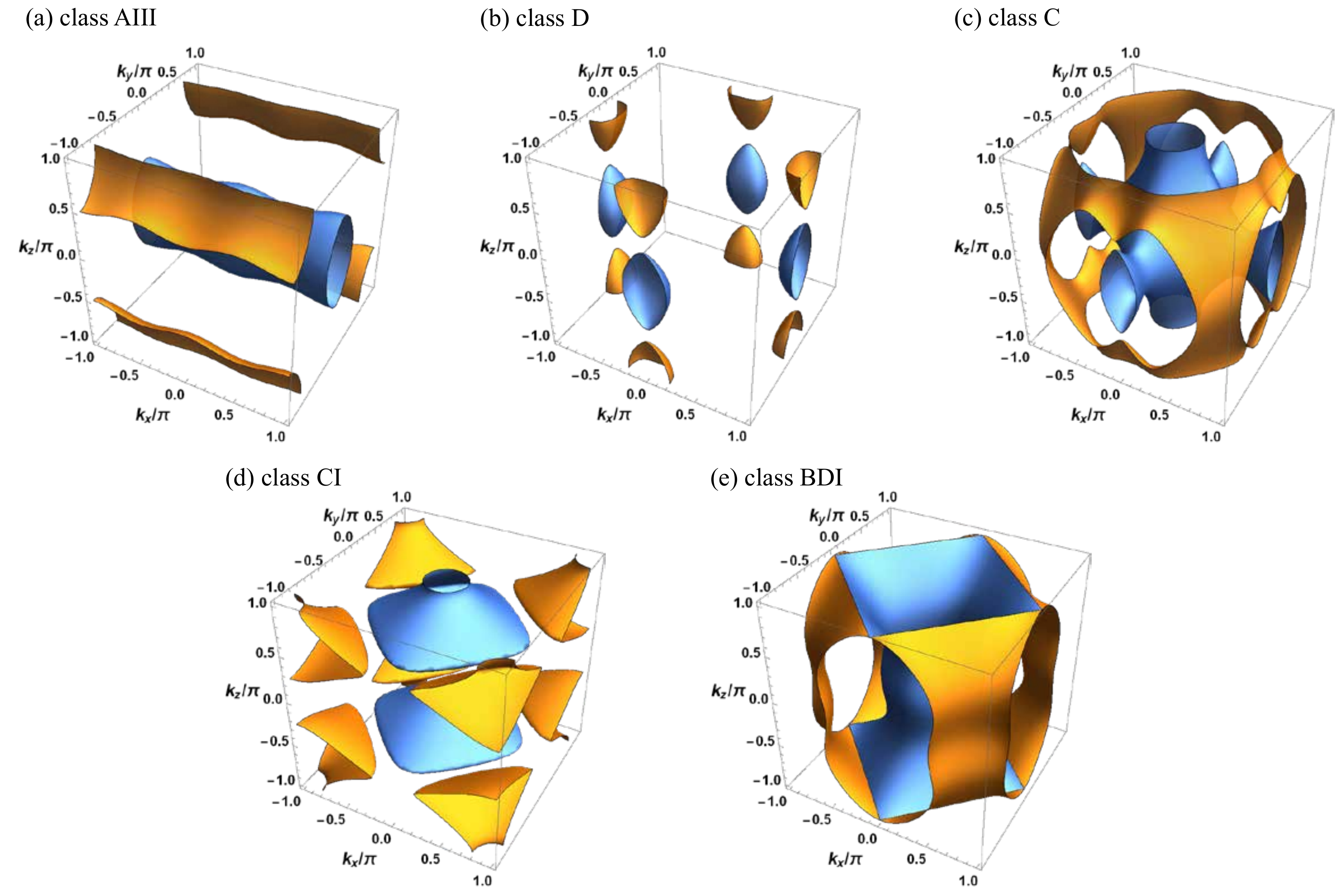}
	\caption{\label{fig1}
Fermi surfaces of the two-bands models in the five non-standard symmetry classes at $E=0$ in the clean limit. The yellow surfaces are of the upper band, and the blue surfaces are of the lower band.}
\end{figure*}
Let us first introduce the following two-orbital tight-binding model on 
the cubic lattice that is shared by all the three unitary models 
studied in this paper:
\begin{align}
	H_0 & \equiv \sum_{{\bm i},{\bm j}} \sum_{d,d^{\prime}=a,b}
     |{\bm i},d\rangle \big[\mathbb{H}_0\big]_{({\bm i},d|{\bm j},d^\prime)} \langle {\bm j},d| \nonumber \\
	&= \sum_{\bm i} \left( \epsilon_{\bm{i}} + \Delta \right) 
	\left( |{\bm i},a\rangle \langle {\bm i},a| - 
	|{\bm i},b\rangle \langle {\bm i}, b| \right)\!\ + \nonumber \\
	& \sum_{\bm i}  \bigg\{ t_{\perp} \Big[ \mathrm{i} 
	\left(|{\bm i}+{\bm e}_x,a\rangle \langle {\bm i}, b| 
    + |{\bm i}+ {\bm e}_x,b\rangle 
	\langle {\bm i}, a|\right) \nonumber \\
	& \ \ \ + \left(|{\bm i}+{\bm e}_y,a\rangle \langle {\bm i},b| - 
	|{\bm i}+{\bm e}_y,b\rangle 
   \langle {\bm i},a|\right)\Big] \nonumber \\
	& \  + t_{\parallel} 
 \left(|{\bm i}+{\bm e}_z,a\rangle \langle {\bm i},a| 
  + |{\bm i}+{\bm e}_z,b\rangle \langle {\bm i},b|\right) 
  + \mathrm{H.c.} \bigg\}. \label{eq:H0}
\end{align}
Here ${\bm i}\equiv (i_x,i_y,i_z)$ specifies the cubic-lattice site; 
${\bm e}_{x}\equiv (1,0,0)$, ${\bm e}_{y}\equiv (0,1,0)$, and 
${\bm e}_{z}\equiv (0,0,1)$. The lattice constant is taken to be the unit length. $a$
and $b$ are the two orbitals: $d,d^{\prime}=a,b$.  
The nearest neighbor (NN) hopping 
within the $x$-$y$ plane is an inter-orbital hopping $t_{\perp}$, 
and the NN hopping along $z$ direction is an intra-orbital hopping $t_{\parallel}$.  
$\pm \Delta$ is an onsite energy for the $a,b$ orbitals respectively, 
and $\epsilon_{\bm{i}}$ is a random potential independently distributed on each lattice site. In numerical simulations, we choose 
the distribution of the random potential to be 
uniform: $\epsilon_{\bm{i}} \in [-W/2,W/2],~ \overline{\epsilon_{\bm{i}}\epsilon_{\bm{j}}}=\delta_{\bm{ij}}W^2/12$.
The tight binding Hamiltonian has the following chiral symmetries (Eqs.~(\ref{p1},\ref{p2})) 
and particle-hole symmetries (Eqs.~(\ref{c1}),(\ref{c2})):
\begin{align}
    \mathbb{P}_1\mathbb{H}_0 \mathbb{P}^{-1}_1 &= - \mathbb{H}_0, \ \ 
    \mathbb{P}_1 = \delta_{{\bm i},{\bm j}}
    (-1)^{i_x+i_z}[\sigma_2]_{d,d^{\prime}}, \label{p1} \\ 
    \mathbb{P}_2\mathbb{H}_0 \mathbb{P}^{-1}_2
    &= - \mathbb{H}_0, \ \ 
    \mathbb{P}_2 = 
    \delta_{{\bm i},{\bm j}}
    (-1)^{i_y+i_z}[\sigma_1]_{d,d^{\prime}}, \label{p2} \\ 
        \mathbb{C}_1\mathbb{H}^T_0 \mathbb{C}^{-1}_1 &= - \mathbb{H}_0, \ \ 
    \mathbb{C}_1 = \delta_{{\bm i},{\bm j}}
    (-1)^{i_z}[\sigma_1]_{d,d^{\prime}}, \label{c1} \\ 
    \mathbb{C}_2\mathbb{H}^T_0 \mathbb{C}^{-1}_2 &= - \mathbb{H}_0, \ \ 
    \mathbb{C}_2 = 
    \delta_{{\bm i},{\bm j}}
    (-1)^{i_x+i_y+i_z}[\sigma_2]_{d,d^{\prime}}, \label{c2}  
\end{align}
with $\mathbb{P}_2 \equiv \mathbb{P}_1\mathbb{C}_1 \mathbb{C}_2$. 
$\sigma_{1,2,3}$ are the Pauli matrices in the $a,b$ orbital space 
and $\sigma_0$ is the unit matrix. The combination of 
particle-hole symmetries and chiral symmetries gives the 
time-reversal symmetries, e.g. $(\mathbb{P}_1\mathbb{C}_1)^T 
\mathbb{H}^T_0 \mathbb{P}_1\mathbb{C}_1 = \mathbb{H}_0$.   
$\mathbb{P}_1\mathbb{P}_2 \equiv  \mathbb{C}_1\mathbb{C}_2=(-1)^{i_x+i_y}
\mathrm{i}\sigma_z$ commutes with $\mathbb{H}_0$, where   
the tight-binding Hamiltonian can be block-diagonalized in a basis of 
the real-valued eigenvectors of $\mathbb{P}_1\mathbb{P}_2$. 
Since $\mathbb{P}_1\mathbb{C}_1$ and $\mathbb{P}_1\mathbb{C}_2$ commute with $\mathbb{P}_1\mathbb{P}_2$,  $\mathbb{P}_1\mathbb{C}_1$ and $\mathbb{P}_1\mathbb{C}_2$ are block-diagonalized in the same 
basis, too. Accordingly, the two blocks of the tight-binding Hamiltonian are 
time-reversal symmetric. Since 
$(\mathbb{P}_1\mathbb{C}_1)^*(\mathbb{P}_1\mathbb{C}_1) = +1$,  
$H_0$ belongs to the orthogonal class. On the basis of such $H_0$, 
we will next introduce other parts of the hopping terms, which 
lead to the three unitary models belonging to the three non-standard symmetry
classes one by one. 

\subsection{Class AIII model}
To break the two particle-hole symmetries, we add the following 
nearest neighbor hoppings within the $x$-$y$ plane;
\begin{align}
    H_{\rm AIII} & \equiv \sum_{{\bm i},{\bm j}} \sum_{d,d^{\prime}=a,b}
     |{\bm i},d\rangle \big[\mathbb{H}_{\rm AIII}\big]_{({\bm i},d|{\bm j},d^\prime)} \langle {\bm j},d^{\prime}| \nonumber \\
     & = H_0 + \sum_{\bm i}  \Big[ t_{1} 
	\left(|{\bm i}+{\bm e}_x,a\rangle \langle {\bm i}, a| 
    - |{\bm i}+ {\bm e}_x,b\rangle 
	\langle {\bm i}, b|\right). \nonumber \\
	&+ t_2 \left(|{\bm i}+{\bm e}_y,a\rangle \langle {\bm i},a| + 
	|{\bm i}+{\bm e}_y,b\rangle 
   \langle {\bm i},b|\right) + \mathrm{H.c.}\Big]. \label{a3}
\end{align}
The model respects the chiral symmetry in Eq.~(\ref{p2}), while the 
two particle-hole symmetries are broken by $t_2$ and $t_1$, respectively. 
We set $\Delta=0.0$, $t_\perp = 0.6$, $t_{\parallel} = 0.4$, and $t_1=t_2=0.5$. 
The Hamiltonian in the clean limit is Fourier-transformed to  
\begin{align}	
	H_{\mathrm{AIII}}(\bm{\mathrm{k}}) =& 2(t_{\parallel}\cos k_z + t_2 \cos k_y) \sigma_0 + (\Delta + 2t_1 \cos k_x) \sigma_3 \nonumber \\
	& -2t_{\perp}(\sin k_x \sigma_1 + \sin k_y \sigma_2),
	\label{eq:Hk_AIII}
\end{align}
where the two bands are particle-hole symmetric under the translation of the momentum by $(0,\pi,\pi)$. The upper (lower) energy band forms an electron (hole) pocket at $E=0$ around the zone boundary (center) axis of $(k_y,k_z)=(\pi,\pi)$ ($(0,0)$) (Fig.~\ref{fig1}(a)).
\subsection{Class D model}
To break the chiral symmetries and one of the two particle-hole symmetries 
(Eq.~(\ref{c2})), we add to $H_0$ the following NN hoppings within the $x$-$y$ plane:
\begin{align}
    H_{\rm D} & \equiv \sum_{{\bm i},{\bm j}} \sum_{d,d^{\prime}=a,b}
     |{\bm i},d\rangle \big[\mathbb{H}_{\rm D}\big]_{({\bm i},d|{\bm j},d^\prime)} \langle {\bm j},d^{\prime}| \nonumber \\
     & = H_0 + \sum_{\bm i}  \Big[ t_{1} 
	\left(|{\bm i}+{\bm e}_x,a\rangle \langle {\bm i}, a| 
    - |{\bm i}+ {\bm e}_x,b\rangle 
	\langle {\bm i}, b|\right). \nonumber \\
	& + t_2 \left(|{\bm i}+{\bm e}_y,a\rangle \langle {\bm i},a| - 
	|{\bm i}+{\bm e}_y,b\rangle 
   \langle {\bm i},b|\right)+ \mathrm{H.c.}\Big]. \label{eq:H_D}
\end{align}
The model respects the particle-hole symmetry in Eq.~(\ref{c1}), while the two 
chiral symmetries are broken by $t_1$ and $t_2$, respectively. 
We set $\Delta=0.1$, $t_\perp = 0.3$, $t_{\parallel} = 0.2$, and $t_1=t_2=0.5$. 
Since $\mathbb{C}^T_1 = \mathbb{C}_1$, the model 
belongs to the symmetry class D. 
The tight-binding model can be regarded as layered Chern insulator models \cite{LiuShang2016,Burkov2011WSM}. In momentum space, the clean-limit 
Hamiltonian takes the following form
\begin{align}	
	H_{\mathrm{D}}(\bm{\mathrm{k}}) &= 2t_{\parallel}\cos k_z \sigma_0 - 2t_{\perp}(\sin k_x \sigma_1 + \sin k_y \sigma_2)\nonumber \\
	& + (\Delta + 2t_1 \cos k_x + 2t_2 \cos k_y) \sigma_3 .
	\label{eq:Hk_D}
\end{align}
A finite $t_{\parallel}$ closes the band gap at $E=0$, where 
electron and hole pockets appears at $k_z=0$ and $\pi$, respectively, as shown in  Fig.~\ref{fig1}(b). 
\subsection{Class C model}
To construct a class C model from Eq.~(\ref{eq:H0}), we 
add to $H_0$ the following nearest neighbor hoppings 
within the $x$-$y$ plane:
\begin{align}
    H_{\rm C} & \equiv \sum_{{\bm i},{\bm j}} \sum_{d,d^{\prime}=a,b}
     |{\bm i},d\rangle \big[\mathbb{H}_{\rm C}\big]_{({\bm i},d|{\bm j},d^\prime)} \langle {\bm j},d^{\prime}| \nonumber \\
     & = H_0 + \sum_{\bm i}  \Big[ t_{1} 
	\left(|{\bm i}+{\bm e}_x,a\rangle \langle {\bm i}, a| 
    + |{\bm i}+ {\bm e}_x,b\rangle 
	\langle {\bm i}, b|\right). \nonumber \\
	&+ t_2 \left(|{\bm i}+{\bm e}_y,a\rangle \langle {\bm i},a| + 
	|{\bm i}+{\bm e}_y,b\rangle 
   \langle {\bm i},b|\right)+ \mathrm{H.c.}\Big]. \label{eq:H_C}
\end{align}
The model respects the particle-hole symmetry Eq.~(\ref{c2}) with 
$\mathbb{C}^T_2=-\mathbb{C}_2$, while finite $t_1$ and $t_2$ break Eq.~(\ref{p2}) and (\ref{p1}), respectively. Therefore the model belongs to symmetry class C. 
In this paper we set $\Delta=0.1$, $t_\perp = 0.6$, $t_{\parallel} = 0.4$, and 
$t_1=t_2=0.5$. In the clean limit, this Hamiltonian is Fourier-transformed to
\begin{align}	
	H_{\mathrm{C}}(\bm{\mathrm{k}}) & =  2(t_{\parallel}\cos k_z + t_1 \cos k_x + t_2 \cos k_y) \sigma_0 + \Delta \sigma_3 \nonumber \\
	& -2t_{\perp}(\sin k_x \sigma_1 + \sin k_y \sigma_2). \label{eq:Hk_C}
\end{align}
The Fermi surface at $E=0$ is shown in Fig.~\ref{fig1}(c). 

The above models respect only one of the four (particle-hole or chiral) symmetries, Eqs.~(\ref{p1}), (\ref{p2}), (\ref{c1}),
or (\ref{c2}).  If there were time reversal symmetry $\mathbb{T}$, $\mathbb{T}\mathbb{H}^T \mathbb{T}^{-1} = \mathbb{H}$,
combining $\mathbb{T}$ and the particle-hole or chiral symmetry should give rise to another chiral or particle-hole symmetry.
Having only one particle-hole or chiral symmetry, therefore, 
means TRS is broken. 
In Appendix \ref{A}, we explicitly demonstrate the broken-time-reversal 
symmetries in the these three models, by showing the 
non-zero Hall conductivity calculated from the Berry phase of Bloch bands.

\section{Orthogonal models with particle-hole or chiral symmetries}
In this chapter, we introduce two-orbital 
tight-binding models that belong to 
two orthogonal classes CI and BDI with chiral and particle-hole symmetries.
The two models are 
real-valued and symmetric, hence chiral and particle-hole symmetries are equivalent. In the 
following two models, the random potential 
$\epsilon_{\boldsymbol{i}}$ takes the same uniform distribution 
as in Eq.~(\ref{eq:H0}). The class CI model is given as follows,
\begin{align}
	H_\mathrm{CI} & \equiv \sum_{\boldsymbol{i},{\boldsymbol j}} \sum_{d,d^{\prime}=a,b}
	|{\boldsymbol i},d\rangle \big[\mathbb{H}_\mathrm{CI}\big]_{({\boldsymbol i},d|{\boldsymbol j},d^\prime)} \langle {\boldsymbol j},d^{\prime}| \nonumber \\
	&= \sum_{\boldsymbol i} \left( \epsilon_{\boldsymbol{i}} + \Delta \right) 
	\left( |{\boldsymbol i},a\rangle \langle {\boldsymbol i},b| + 
	|{\boldsymbol i},b\rangle \langle {\boldsymbol i}, a| \right)\!\  \nonumber \\
	& +\sum_{\boldsymbol i}\bigg\{\sum_{\mu=x,y} t_{\perp}
	\big(|{\boldsymbol i}+{\boldsymbol e}_{\mu},a\rangle \langle {\boldsymbol i}, a| +
	|{\boldsymbol i}+{\boldsymbol e}_{\mu},b\rangle \langle {\boldsymbol i}, b| \big) \nonumber \\
	& + t_{\parallel}
	\big(|{\boldsymbol i}+{\boldsymbol e}_z,a\rangle \langle {\boldsymbol i},a| 
	- |{\boldsymbol i}+{\boldsymbol e}_z,b\rangle \langle {\boldsymbol i},b|\big) \nonumber \\
	& + t_{\parallel}^{\prime} \big(|{\boldsymbol i}+{\boldsymbol e}_z,a\rangle \langle {\boldsymbol i},b| + 
	|{\boldsymbol i}+{\boldsymbol e}_z,b\rangle \langle {\boldsymbol i},a|\big)
	+ \mathrm{H.c.} \bigg\}. \label{eq:H_CI}
\end{align}
It has only the following particle-hole (chiral) symmetry,
\begin{align}
    \mathbb{C}_3\mathbb{H}^T_0 \mathbb{C}^{-1}_3 &= - \mathbb{H}_0, \ \ 
    \mathbb{C}_3 = 
    \delta_{{\bm i},{\bm j}}
    (-1)^{i_x+i_y}[\sigma_2]_{d,d^{\prime}}, \label{c3}
\end{align}
with $\mathbb{C}_3^T = - \mathbb{C}_3$. Note that there is no other particle-hole symmetries. Namely, 
$\mathbb{C}_3$ must be diagonal with respect to the cubic-lattice site index, because of the on-site random terms. From the on-site term, it has to take a form of either $\sigma_2$ or $\sigma_3$
in the orbital space. From the nearest-neighbor hoppings within the $x$-$y$ plane,
it comes with a U(1) phase of $(-1)^{i_x+i_y} e^{i\theta(i_z)}$.
From the hopping along $z$, such $\mathbb{C}_3$ is uniquely determined by $(-1)^{i_x+i_y} \sigma_2$, hence
the model belongs to the symmetry class CI.
In this paper, we take $\Delta=t_{\perp}=t_{\parallel}=1,\ t_{\parallel}^{\prime}=2$.
The Fermi surface at $E=0$ is shown in Fig.~{\ref{fig1}}(d).

The class BDI model is given by 
\begin{align}
	H_\mathrm{BDI} & \equiv \sum_{\boldsymbol{i},{\boldsymbol j}} \sum_{d,d^{\prime}=a,b}
	|{\boldsymbol i},d\rangle \big[\mathbb{H}_\mathrm{BDI}\big]_{({\boldsymbol i},d|{\boldsymbol j},d^\prime)} \langle {\boldsymbol j},d^{\prime}| \nonumber \\
	&= \sum_{\boldsymbol i} \left( \epsilon_{\boldsymbol{i}} + \Delta \right) 
	\left( |{\boldsymbol i},a\rangle \langle {\boldsymbol i},a| - 
	|{\boldsymbol i},b\rangle \langle {\boldsymbol i}, b| \right)\!\  \nonumber \\
	& +\sum_{\boldsymbol i}\bigg\{\sum_{\mu=x,y} t_{\perp}
	\big(|{\boldsymbol i}+{\boldsymbol e}_{\mu},a\rangle \langle {\boldsymbol i}, a| +
	|{\boldsymbol i}+{\boldsymbol e}_{\mu},b\rangle \langle {\boldsymbol i}, b| \big) \nonumber \\
	& + t_{\parallel}\big(|{\boldsymbol i}+{\boldsymbol e}_z,a\rangle \langle {\boldsymbol i},a| - 
	|{\boldsymbol i}+{\boldsymbol e}_z,b\rangle 
	\langle {\boldsymbol i},b|\big) \nonumber \\
	& + t_{\parallel}^{\prime}
	\big(|{\boldsymbol i}+{\boldsymbol e}_z,a\rangle \langle {\boldsymbol i},b| 
	- |{\boldsymbol i}+{\boldsymbol e}_z,b\rangle \langle {\boldsymbol i},a|\big) 
	+ \mathrm{H.c.} \bigg\}. \label{eq:H_BDI}
\end{align}
It has only the following 
chiral (particle-hole) symmetry;
\begin{align}
    \mathbb{P}_3\mathbb{H}_{\rm BDI} \mathbb{P}^{-1}_3 &= - \mathbb{H}_{\rm BDI}, \ \ 
    \mathbb{P}_3 = 
    \delta_{{\bm i},{\bm j}}
    (-1)^{i_x+i_y}[\sigma_1]_{d,d^{\prime}}.
\end{align}
with $\mathbb{P}_3^T = \mathbb{P}_3$. Namely, from the first two terms of Eq.~(\ref{eq:H_BDI}), $\mathbb{P}_3$ must take a form of either $(-1)^{i_x+i_y} e^{i\theta(i_z)} \sigma_1$ or $(-1)^{i_x+i_y} e^{i\theta(i_z)} \sigma_2$. The last two terms uniquely determine $\mathbb{P}_3$ to be  $(-1)^{i_x+i_y}  \sigma_1$; 
the model belongs to the symmetry class BDI. 
Here we take 
$\Delta=t_{\perp}=t_{\parallel}^{\prime}=1,\ t_{\parallel}=0.5$.
The Fermi surface at $E=0$ is shown in Fig.~{\ref{fig1}}(e) 

We emphasize that the BdG symmetry classes describe 
the localization of Bogoliubov quasiparticle wavefunctions in 
superconductors~\cite{AZ1997PRB}, where disordered single-particle 
Hamiltonian is given by a mixed basis in particle and hole space. In 
the class C, D and CI models introduced above, the particle and hole 
degrees of freedom are undertaken by the two orbitals. Thereby, the
concept of a finite area of Fermi surface applies to 
single-particle wavefunctions of 
these two-orbitals models. We expect the 
critical behaviors of the Anderson transition of the 
single-particle wavefunctions are the same as those 
of the superconducting quasiparticle wavefunctions.

\section{Method and Result}
\begin{figure*}
	\includegraphics[width=6.5in]{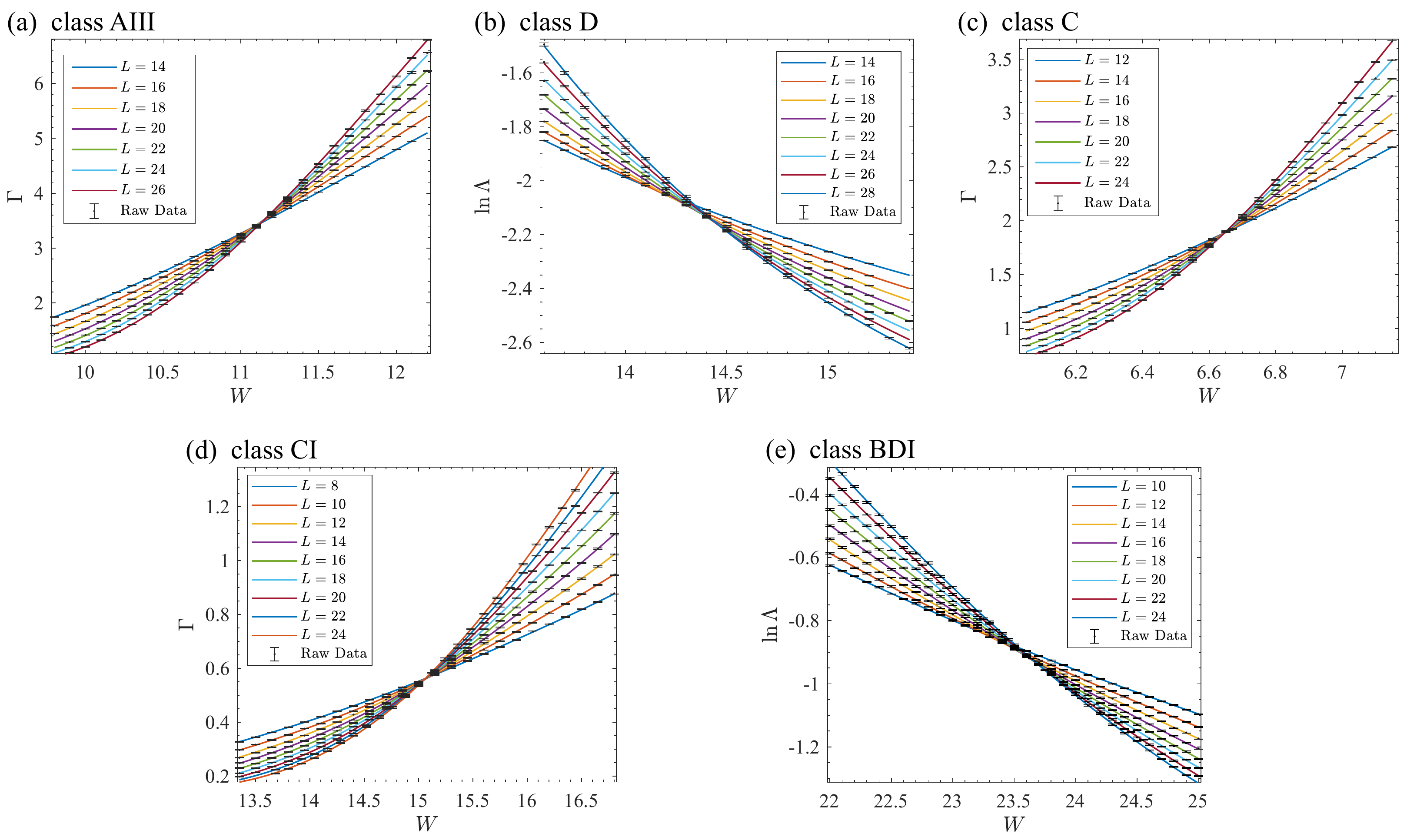}
	\caption{\label{fig2} Plots of normalized localization length (or its inverse or its logarithm) of the five models near the critical point of the Anderson transition for different system size in the transverse direction (see text). $\Lambda$ is the dimensionless localization length and $\Gamma=\Lambda^{-1}$. Black dots are numerical data points with error bar and colored curves are the fitting results with the largest GOF in TABLE~\ref{tab1}. The quasi-one-dimensional samples are typically of length $L_z = 10^6 - 10^7$ to ensure a 0.2 \%  precision of each data point.}
\end{figure*}

\begin{figure*}
	\includegraphics[width=4.5in]{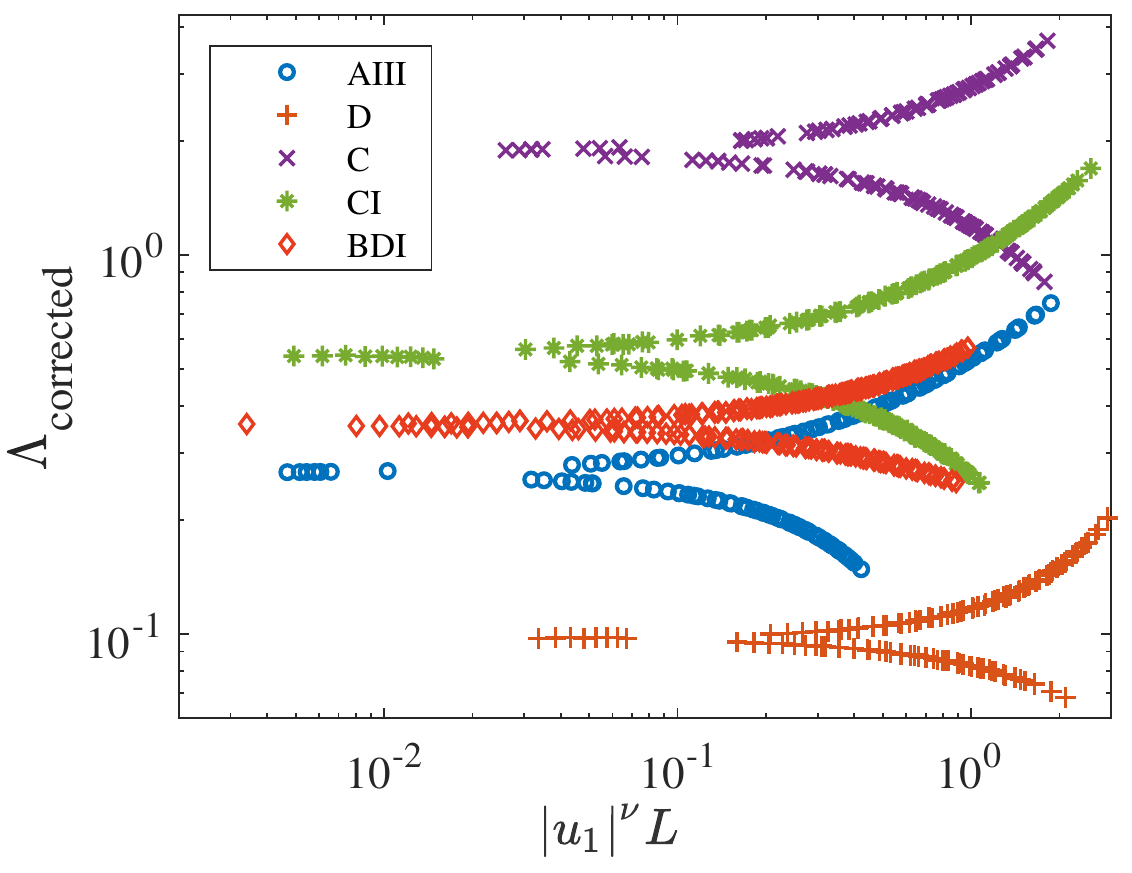}
	\caption{\label{fig3} The single-parameter scaling function of the normalized localized length for the five symmetry classes. $\Lambda_{\mathrm{corrected}}$ is the numerical data $\Lambda$ subtracted by all terms of irrelevant scaling variable in Eq.~(\ref{eq:expand}). The upper branch is for the delocalized phase and the lower branch is for the insulator phase.	We note that $\Lambda_c$ depends on the anisotropy of the model.}
\end{figure*}
The scaling theory is at the heart of studying quantum criticality of the 
Anderson localization. On approaching the critical point, various physical 
quantities diverge according to a power law with the universal critical exponent $\nu$.
One of them is the correlation length $\xi \sim |x-x_c|^{-\nu}$, which in a 
localized phase, characterises a decay length of a density-density correlation 
function in disordered medium. Here $x$ represents the coordinate in a 
certain parameter space of the system and $x_c$ stands for the critical point.
To determine the critical exponent, we calculate a quasi-one-dimensional (Q1D)
localization length $\lambda$ of a single-particle wavefunction amplitude, 
using transfer matrix method \cite{MacKinnon81,MacKinnon83}. To be specific, 
we consider the Q1D geometry of the 3D 
cubic lattice ($L_x\times L_y \times L_z$) with a cross-section 
$L_x=L_y=L \ll L_z$. In such geometry, all the eigenstates of the 
disordered Hamiltonian are localized in $z$ direction, 
$|\psi|\sim\exp(-|z-z_0|/\lambda)$, with the localization length $\lambda$. 
The scaling argument around the critical point suggests that the 
localization length normalized by the linear dimension of the cross-section, 
$\Lambda\equiv \lambda/L$ or $\Gamma=\Lambda^{-1}$, must be 
scale-invariant (independent of $L$) at the critical point (see Fig.~\ref{fig2}).  
Namely, numerical data of $\Lambda$ for different 
cross-sectional size $L$ must fit in so-called the one-parameter 
scaling function near the transition point~\cite{MacKinnon81,MacKinnon83,Pichard81},
\begin{align}
	\Lambda(W,L)=f(L/\xi(W))
	\label{eq:onescaling}
\end{align}
where $f(x)$ is a universal scaling function for the normalized localization 
length. $\xi(W)$ is the correlation length that depends on disorder strength: 
$\xi(W) \sim |W-W_c|^{-\nu}$. In practical calculations, one often encounters 
deviations of the data sets from the single-parameter scaling form. 
The deviations are attributed to finite-size effects associated with 
irrelevant scaling variables around the critical point in the framework 
of a renormalization group (RG) theory of general critical phenomena. In the RG 
theory, generic continuous phase transition is characterized by a saddle-point 
fixed point of RG equations for a certain effective theory with multiple 
system parameters. When high-energy degrees of freedom are integrated out, 
low-energy system parameters run away from the 
fixed point only along one direction in the parameter space, while 
they flow into the fixed point along all the other directions. 
The unstable direction around the fixed point is characterized by relevant scaling variable, 
while its complementary directions are characterized by a number of irrelevant 
scaling variables. Scaling dimension of the relevant scaling variable is 
positive and it defines the universal critical exponent, $1/\nu$. Scaling 
dimensions of the irrelevant scaling variables are all negative, where 
the irrelevant variable with the largest negative scaling dimension is the 
least irrelevant. Empirically, we attribute the deviation of the numerical 
data from the single-parameter scaling to the least irrelevant scaling variable~\cite{SlevinTomi1999,Slevin14}.  
The scaling argument that includes the effect of the least irrelevant scaling
variable modifies the single-parameter scaling function form into  
\begin{align}
	\Lambda(W,L) = F\left(u_1(w) L^{1/\nu},u_2(w) L^{-y}\right),
	\label{eq:twoscaling}
\end{align}
where $w=(W-W_c)/W_c$, and $1/\nu>0$ and $-y<0$ are scaling dimensions 
of the relevant scaling variable $u_1$ and the least irrelevant scaling 
variable $u_2$, respectively. The omission of the irrelevant scaling variables 
with smaller negative scaling dimensions are justified a posterior by
large $y$ obtained from the fitting (see TABLE~\ref{tab1}).  

To obtain the exponents $\nu$, $y$ as well as the scaling function $f(x) 
\equiv F(x^{1/\nu},0)$, we expand the scaling function $F$ as Taylor series 
of its arguments
\begin{align}
	\Lambda = \sum_{i=0}^{n_1} \sum_{j=0}^{n_2} a_{ij} \left(u_1 L^{1/\nu}\right)^i \left(u_2 L^{-y}\right)^j.
	\label{eq:expand}
\end{align}
When the system is close to the critical point, the scaling variables can be 
also expanded in small $w$
\begin{align}
	u_1 = \sum_{k=1}^{m_1} b_{k} w^k,\quad u_2 = \sum_{k=0}^{m_2} c_{k} w^k.
\end{align}
To fix the ambiguity of fitting, we at $a_{1,0}=a_{0,1}=1$~\cite{SlevinTomi1999,Slevin14}.
Having the polynomial form of the scaling function $F$ in $w$, 
we use the least-squares fitting procedure and minimize $\chi^2$ defined as
follows,
\begin{align}
	\chi^2 = \sum_{i=1}^{N_D} \frac{(\Lambda_i - F_i)^2}{\sigma_i^2}.
\end{align}
Here $\Lambda_i,\,\sigma_i,\,F_i$ are the numerical mean value of 
$\Lambda$, numerical standard error of $\Lambda$, 
and a fitting value of $\Lambda$ from the polynomial fitting function. 
They are given at every 
data point specified by $L$ and $W$ with the data point index $i=1,\cdots,N_D$. 
The fitting results give estimate of $\nu,\, y, \, W_c, \, \Lambda_c$ and 
expansion coefficients $\{a_{ij}\},\,\{b_k\},\,\{c_k\}$.  
We choose $n_1=3$, $n_2=1$ and perform the polynomial fitting 
for different values of $m_1$ and $m_2$. Results are shown
in TABLE \ref{tab1}.  They are stable 
against changes of the expansion order $m_1$, $m_2$, as well as the range of system sizes. 
The goodness of fit (GOF) is a probability that the $N_D$ data points 
sampled from the fitting function would give a $\chi^2$ larger than 
the minimized value $\chi^2_{\text{min}}$ of the fitting \cite{DataReduction}. 
We use $\Gamma$ as data inputs for the evaluations in the class AIII, C 
and CI models, and $\ln\Lambda$ for the class D and BDI models, in order to 
get better fitting qualities. The resulting critical exponents of 
the five symmetry classes are
\begin{align}
	\nu_{\text{AIII}}\ =&\ 1.06 \pm 0.02, \nonumber\\
	\nu_{\text{D}}\ =&\ 0.87 \pm 0.03, \nonumber\\
	\nu_{\text{C}}\ =&\ 0.996 \pm 0.012,\nonumber\\
	\nu_{\text{CI}}\ =&\ 1.17 \pm 0.02,\nonumber\\
	\nu_{\text{BDI}}\ =&\ 1.12 \pm 0.06.
\end{align}
The numbers after $\pm$ sign are the standard deviation 
determined from 1000 fittings of synthetic data sets. 

The values of the evaluated critical exponents 
in the three  non-standard symmetry classes with broken time-reversal symmetry 
are distinct from each other and also 
from previously evaluated critical exponents in other symmetry classes. 
The symmetry class C in 3D has been studied by network model of both quantum 
systems \cite{Kagalovsky3dclassC} and the classical 
counterpart \cite{Chalker3dclassC}. 
The evaluated critical exponent 
in Ref.~\cite{Chalker3dclassC} is 
$\nu=0.9985\pm0.0015$, which is 
in good agreement with our result. 
The symmetry class CI in 3D has also been studied with a lattice 
model \cite{Luo2020PRB} as $\nu=1.16\pm 0.02$, also in 
good agreement with our result. These consistencies 
demonstrate that the universality class 
of the Anderson transitions are free from details of 
the models, and it is determined only by the symmetry 
of the random matrices and the spatial dimension.

In Fig.~\ref{fig3}, we show the scaling functions of 
these five non-standard symmetry classes.
We define $\Lambda_{\text{corrected}}$ from raw numerical data
subtracted by the contributions of the irrelevant scaling variable 
in the fitting function $F$.
$\Lambda_{\text{corrected}}$ should obey 
the single-parameter scaling form in Eq.~(\ref{eq:onescaling}) around 
the critical point, with the scaling argument $L/\xi\sim L |u_1|^{\nu}$.
When plotted as a function of $L |u_1|^{\nu}$, $\Lambda_{\text{corrected}}$
all collapse onto two branches of the scaling functions, 
where the upper (lower) branch corresponds to the metallic (localized) phase regime. 

\begin{table*}
	\caption{\label{tab1} Finite-size scaling analyses of the normalized localization length around the Anderson transition in the five non-standard symmetry classes. The scaling function of the normalized localization length is expanded in power of $w \equiv (W-W_c)/W_c$ (see text). The expansion orders $n_1=3,\ n_2=1$ are fixed, while $m_1,m_2$ are variables. The values inside the square brackets are 95\% confidence intervals from 1000 Monte Carlo simulations. Only fitting samples with good of fit (GOF) greater than 0.1 are shown.
	We note that our model is anisotropic, and $\Lambda_c$ depends on the anisotropy.}
	\flushleft
	(a) 3D class AIII
	\begin{ruledtabular}
		\begin{tabular}{cccccccc}
			$m_1$ & $m_2$ & $L$ & GOF & $ W_c$ & $\nu$ & $y$ & $\Lambda_c$ \\
			\hline
			2 & 0 & 12-26 & 0.308 & 11.247[11.226, 11.272] & 1.059[1.022, 1.100] & 0.814[0.585, 1.042] & 0.248[0.234, 0.258] \\
			2 & 0 & 14-26 & 0.475 & 11.223[11.201, 11.252] & 1.071[1.033, 1.117] & 1.326[0.975, 1.695] & 0.262[0.251, 0.269] \\
			\hline
			3 & 0 & 12-26 & 0.290 & 11.255[11.234, 11.278] & 1.069[1.030, 1.110] & 0.726[0.527, 0.934] & 0.244[0.228, 0.254] \\
			3 & 0 & 14-26 & 0.479 & 11.213[11.193, 11.238] & 1.055[1.013, 1.097] & 1.572[1.144, 2.010] & 0.266[0.257, 0.272] \\
		\end{tabular}
	\end{ruledtabular}
	\flushleft
	(b) 3D class D
	\begin{ruledtabular}
		\begin{tabular}{cccccccc}
			$m_1$ & $m_2$ & $L$ & GOF & $ W_c$ & $\nu$ & $y$ & $\Lambda_c$ \\
			\hline
			2 & 0 & 14-26 & 0.155 & 14.55[14.51, 14.61] & 0.862[0.800, 0.903] & 1.589[1.108, 2.069] & 0.105[0.097, 0.109] \\
			2 & 0 & 14-28 & 0.173 & 14.61[14.57, 14.67] & 0.870[0.793, 0.928] & 1.126[0.811, 1.444] & 0.097[0.092, 0.103] \\
			\hline
			3 & 0 & 12-26 & 0.147 & 14.55[14.50,14.61 ] & 0.869[0.804, 0.915] & 1.624[1.156, 2.113] & 0.105[0.098, 0.109] \\
			3 & 0 & 14-28 & 0.169 & 14.60[14.55, 14.66] & 0.904[0.827, 0.967] & 1.239[0.889, 1.587] & 0.099[0.091, 0.104] \\
		\end{tabular}
		\flushleft
		(c) 3D class C
		\begin{ruledtabular}
			\begin{tabular}{cccccccc}
				$m_1$ & $m_2$ & $L$ & GOF & $ W_c$ & $\nu$ & $y$ & $\Lambda_c$ \\
				\hline
				3 & 1 & 12-24 & 0.313 & 6.641[6.637, 6.644] & 0.9957[0.9711, 1.0184] & 1.88[1.64, 2.38] & 0.535[0.531, 0.538] \\
				4 & 0 & 12-24 & 0.287 & 6.642[6.639, 6.645] & 0.9972[0.9800, 1.0111] & 2.86[2.43, 3.34] & 0.533[0.531, 0.536] \\
				4 & 1 & 12-24 & 0.295 & 6.641[6.639, 6.646] & 0.9967[0.9738, 1.0207] & 2.00[1.46, 2.57] & 0.534[0.531, 0.538]
			\end{tabular}
		\end{ruledtabular}
	\end{ruledtabular}
	\flushleft
	(d) 3D class CI
	\begin{ruledtabular}
		\begin{tabular}{cccccccc}
			$m_1$ & $m_2$ & $L$ & GOF & $ W_c$ & $\nu$ & $y$ & $\Lambda_c$ \\
			\hline
			2 & 1 & 8-24 & 0.256 & 15.030[15.018, 15.043] & 1.149[1.119, 1.177] & 1.25[1.14, 1.62] & 1.823[1.805, 1.840] \\
			2 & 1 & 10-24 & 0.312 & 15.028[15.009, 15.048] & 1.155[1.096, 1.187] & 1.35[1.23, 1.77] & 1.831[1.798, 1.851]\\
			\hline
			3 & 1 & 8-24 & 0.579 & 15.027[15.013, 15.041] & 1.167[1.132, 1.198] & 1.11[1.02, 1.49] & 1.831[1.807, 1.850] \\
			3 & 1 & 10-24 & 0.533 & 15.025[15.005, 15.047] & 1.174[1.128, 1.215] & 1.21[1.08, 1.76] & 1.835[1.801, 1.864]
		\end{tabular}
	\end{ruledtabular}
	\flushleft
	(e) 3D class BDI
	\begin{ruledtabular}
		\begin{tabular}{cccccccc}
			$m_1$ & $m_2$ & $L$ & GOF & $ W_c$ & $\nu$ & $y$ & $\Lambda_c$ \\
			\hline
			2 & 1 & 10-24 & 0.522 & 23.859[23.803, 23.923] & 1.104[0.940, 1.179] & 1.02[0.77, 1.34] & 0.356[0.338, 0.369]\\
			2 & 1 & 12-26 & 0.311 & 23.895[23.857, 24.000] & 1.116[0.915, 1.201] & 1.04[0.845 1.24] & 0.351[0.337, 0.360]\\
			\hline
			3 & 1 & 10-24 & 0.548 & 23.859[23.756, 23.928] & 1.119[0.973, 1.207] & 1.01[0.76, 1.82] & 0.356[0.338, 0.380] \\
			3 & 1 & 12-26 & 0.325 & 23.896[23.856, 23.949] & 1.115[0.903, 1.241] & 1.04[0.85, 1.22] & 0.351[0.337, 0.360]
		\end{tabular}
	\end{ruledtabular}
\end{table*}

\section{Summary and discussion}
In conclusion, we have clarified the  universal 
critical exponents and scaling functions associated with 
the Q1D localization length of 
the 3D conventional localization-delocalization 
transition in three non-standard 
unitary symmetry classes (the chiral unitary classes AIII, 
the BdG class D and C), and in two non-standard orthogonal classes  
(the chiral orthogonal class BDI, the BdG class CI). 

We introduced two-orbital tight-binding models on the cubic lattice 
that belong to these symmetry classes. In these models, 
the particle-hole degrees of freedom in usual BdG Hamiltonians for
superconductors~\cite{AZ1997PRB} and the sublattice degrees of freedom in usual 
chiral symmetry~\cite{Gade91,verbaarschot93,verbaarschot94,Gade93,slevin93} 
are overtaken by an artificial orbital 
degrees of freedom. One of the advantages of these tight-binding models 
is that all of them have finite area of Fermi surfaces in the 
clean limit. The presence of the finite Fermi surfaces 
in the clean limit guarantees the presence of 
3D Anderson transitions at finite disorder strength. Using 
comprehensive numerical analyses, we (re)determine the critical exponents 
of the Anderson transition in these five non-standard symmetry classes. 
The exponents of the 3D class AIII and  
D are evaluated in this paper for the first time. The exponents 
of class C and CI are consistent with value in previous literature. 

On the contrary, our exponent of the class BDI model is significantly 
different from a previous evaluation in the 
nodal Dirac semimetal (NDS) model~\cite{Luo2020PRB}. The 
disordered NDS model has two topologically different types of 
delocalization-localization 
transitions in its phase diagram. One is a `topological' phase 
transition line between diffusive metal (DM) phase and topological 
insulator (TI) phase with 1D topological winding number 
in the BDI class. The other is a `conventional' phase transition 
line between the DM phase and ordinary (topologically trivial) band 
insulator or Anderson localized phase. The previous evaluation 
of the critical exponents at these two 
types of the phase transition lines are consistent with each other. 
On the one hand, the zero-energy density of state (DOS) suggests that the 
dynamical exponent at these two transitions could be different;  
the zero-energy DOS takes a finite constant value in the former `topological' 
phase transition line, while the DOS shows a weak singularity at the 
zero energy around the `conventional' phase transition 
line~\cite{Luo2020PRB}. 
The singularity of the zero-energy DOS indicates a possible 
deviation of the dynamical exponent from the spatial dimension. 
It is an interesting open issue to study how the zero-energy DOS behaves 
at the Anderson transition point in the present BDI model. Such 
information could provide comprehensive understanding on the 3D 
Anderson transition in the BDI class. 

Recently, the critical exponents and scaling functions in the non-standard 
symmetry classes acquire a lot of research interests from a view point 
of conventional and unconventional delocalization-localization transitions 
in non-Hermitian disordered systems. A previous dimensional regularization 
study of the non-linear sigma models in $d=2+\epsilon$ shows that 
the $\beta$ function in the three chiral symmetry classes is identical to 
zero~\cite{wegner89,Gade91,Gade93}. The vanishing $\beta$ function implies 
an unusual nature of 
delocalization-localization transition in 
these chiral symmetry classes~\cite{konig12}. 
Our evaluation of the universal critical properties provide 
solid information to these research community and will shed a new 
light in future experiments on quasiparticle heat transports in 
superconductors as well as the localization phenomena in 
random dissipative systems.

\begin{acknowledgments}
T. W. and R. S. thanks the fruitful discussion and correspondence 
with Xunlong Luo. T. O. was supported by JSPS KAKENHI Grants 19H00658. T. W. and 
R. S. was supported by the National Basic Research Programs of China 
(No. 2019YFA0308401) and by National Natural Science Foundation of 
China (No.11674011 and No. 12074008). 
\end{acknowledgments}

\appendix
\section{Berry Curvature and Time-reversal Symmetry}
\label{A}
$H$ is time-reversal invariant, when it commutes with
the time reversal operator $\Theta$,
\begin{align}
	[\Theta,H] = 0, \quad \Theta=\mathbb{T}\cdot K\,,
	\label{eq:A1}
\end{align}
where $K$ is the complex conjugation and $\mathbb{T}$ a unitary operator.
Since the tight-binding models studied in this paper contain on-site 
random potentials, the unitary transformation $\mathbb{T}$ must be diagonal 
with respect to the lattice sites. To prove the absence of such $\mathbb{T}$ 
in the class AIII and C models, we calculate the Hall conductivity 
at $E=0$ in the clean limit. 

Two-bands Hamiltonians in the momentum space are written 
in the following form, 
\begin{align}
	H(\bm{\mathrm{k}}) = E_0(\bm{\mathrm{k}}) \sigma_0 + 
	E_1(\bm{\mathrm{k}}) \bm{n}({\bm k})\cdot \bm{\sigma}.
\end{align}
with a three-component unit vector ${\bm n}(\bm k)$ and the 
two by two Pauli matrices ${\bm \sigma}$. 
For the 2 by 2 Hamiltonian, the two 
energy bands and the 
Berry curvature for each band are given by 
\begin{align}
E_{\pm}(\bm{\mathrm{k}}) &= 
E_0 (\bm{\mathrm{k}}) \pm E_1 (\bm{\mathrm{k}}),  \nonumber \\
\Omega_{xy}^{\pm}(\bm{\mathrm{k}}) &= \pm \left(\frac{\partial \bm{n}}{\partial k_x}\times \frac{\partial \bm{n}}{\partial k_y}\right) \cdot \bm{n}. 
\end{align}
In terms of this, the Berry curvature of the class AIII model is
calculated from Eq.~(\ref{eq:Hk_AIII}),
\begin{align}
	\Omega_{xy}^{\pm}(\bm{\mathrm{k}}) = \pm \frac{4t_{\perp}^2}{E_1^3}(\Delta \cos k_x +2t)\cos k_y,
\end{align}
with $E_1(\bm{\mathrm{k}})\equiv[(\Delta+2t\cos k_x)^2 +4t_{\perp}^2(\sin^2 k_x + \sin^2 k_y)]^{1/2}$. 
The Berry curvature of the class D model is calculated 
from Eq.~(\ref{eq:Hk_D}),
\begin{align}
	\Omega_{xy}^{\pm}(\bm{\mathrm{k}}) = \pm \frac{4t_{\perp}^2}{E_1^3}\left[\Delta\cos k_x \cos k_y+2t(\cos k_x + \cos k_y)\right],
\end{align}
with $E_1(\mathbf{k})\equiv [(\Delta+2t\cos k_x + 2t\cos k_y)^2 +4t_{\perp}^2(\sin^2 k_x + \sin^2 k_y)]^{1/2}$. Likewise, 
the Berry curvature of the class C model is calculated 
from Eq.~(\ref{eq:Hk_C}), 
\begin{align}
	\Omega_{xy}^{\pm}(\bm{\mathrm{k}}) = \pm \frac{4t_{\perp}^2}{E_1^3}\Delta\cos k_x \cos k_y,
\end{align}
with $E_1(\bm{\mathrm{k}}) \equiv [\Delta^2 +4t_{\perp}^2(\sin^2 k_x + \sin^2 k_y)]^{1/2}$.

An integral of the Berry curvature over the occupied bands ($E_{\pm}<0$) is 
nothing but the Hall conductivity at $E=0$ at certain $k_z$ \cite{TKNN1982},
\begin{align}
	\sigma_{xy}(k_z) \equiv &\int_{E_{+}<0} \mathrm{d}^2 k \!\  \Omega_{xy}^{+}(\bm{\mathrm{k}}) 
	+ \int_{E_{-}<0}  \mathrm{d}^2 k \!\ \Omega_{xy}^{-}(\bm{\mathrm{k}}).
	\label{eq:sigmaxy}
\end{align}
%\textcolor{blue}{
Throughout this paper we take lattice constant to be one, and the first Brillouin zone is $k_{x,y,z} \in [-\pi,\pi]$. By integrating over $k_z$, one gets the total Hall conductivity $\sigma_{xy}=\int_{-\pi}^\pi \mathrm{d} k_z \sigma_{xy}(k_z)$. 
A non-zero Hall conductivity indicates the broken time-reversal symmetry. 
With the chosen parameters for numerical simulations of the three model, 
we get $\sigma_{xy}/8\pi^2\approx 0.04$ for the class AIII model, $\sigma_{xy}/8\pi^2\approx 0.73$ 
for the class D model, and $\sigma_{xy}/8\pi^2\approx 0.58$ for the class C model.

\bibliography{3dRef}

\end{document}